\documentclass[preprint,prd,letterpaper,nofootinbib,showpacs]{revtex4}
\usepackage{amsmath,amssymb,graphicx}
\usepackage{dcolumn}
\usepackage{bm}
\newcommand{\ds}{\displaystyle}
\newcommand{\pslash}{\bar{p}\llap/}
\newcommand{\ppslash}{p\llap/}
\newcommand{\kslash}{\bar{k}\llap/}
\newcommand{\bgamma}{\bar{\gamma}}
\newcommand{\ssz}{\scriptsize}

\newcommand{\nn}{\nonumber\\}

\newcommand{\ben}{\begin{displaymath}}
\newcommand{\een}{\end{displaymath}}
\newcommand{\be}{\begin{equation}}
\newcommand{\ee}{\end{equation}}
\newcommand{\bea}{\begin{eqnarray}}
\newcommand{\eea}{\end{eqnarray}}

\newcommand{\bp}{\bar{p}}
\newcommand{\bk}{\bar{k}}
\newcommand{\bPi}{\bar{\Pi}}
\newcommand{\eqn}[1]{\label{#1}}
\newcommand{\eq}[1]{Eq.~(\ref{#1})}

\begin{document}
\title{Comment on "Light-front Schwinger model at finite temperature"}
\author{B. Blankleider}\email{boris.blankleider@flinders.edu.au}
\author{A. N. Kvinikhidze}\altaffiliation[On leave from ]{The Mathematical 
Institute of Georgian Academy of Sciences, Tbilisi, Georgia}\email{sasha.kvinikhidze@flinders.edu.au}
\affiliation{Department of Physics, Flinders University, Bedford Park, SA 5042, Australia}
\date{\today}
\begin{abstract}

In a recent paper by A. Das and X. Zhou [Phys.\ Rev.\ D {\bf 68}, 065017 (2003)] it is claimed  that explicit evaluation of the thermal photon self-energy in the Schwinger model gives off-shell thermal Green functions that are different in light-front and conventional quantizations. We show that the claimed difference originates from an erroneous simplification of the fermion propagator used in the light-front calculation.

\end{abstract}
 
\maketitle
In a recent paper by A.\ Das and X.\ Zhou \cite{zhou} it is claimed  that 
explicit evaluation of the finite temperature photon self-energy in the Schwinger model shows  that the photon off-shell thermal Green function is different in light front (LF) and conventional quantizations. It is also claimed that this could be a counterexample to the general proof of equivalence given by us in Ref.\ \cite{blank}. Here we show that the calculated difference obtained in Ref.\ \cite{zhou} is due to an erroneous simplification of the expression for the zero mass fermion propagator, and in no way contradicts the proof of equivalence given in Ref.\ \cite{blank}.

As is stated in Ref.\ \cite{blank}, we have given a proof of equivalence between 
LF and conventional thermal field theories for those models which are equivalent at zero temperature. It is clear, therefore, that this does not imply  that all LF
Green functions are identical to their conventional counterparts, as this statement is not even correct
at zero temperature (see for example Refs.\ \cite{chang1,chang2,yan,yan2}). Yet any 
difference between thermal Green functions can always
be traced to a difference at zero temperature; in other words, what we have shown is that temperature is never the origin of any non-equivalence.

The difference in the finite temperature off-mass-shell photon self-energy calculated
in Ref.\ \cite{zhou}, however, has a totally different origin. In Ref.\ \cite{zhou} the photon self-energy  is calculated in the real time formalism. The LF real time fermion propagator derived in Ref.\ \cite{zhou} is given by their Eq.\ (33), and in the zero mass limit is given by\footnote{Ref.\ \cite{zhou} uses variables defined as $\bp^\mu =(\sqrt{2}p^+,p^1)$ and $\bp_\mu = (p^0,-\sqrt{2}p^+)$ where $p^+=\frac{1}{\sqrt{2}}(p^0+p^1)$.
}
\be
iS_{++}(\bp)= \bar{\gamma}^\mu \bp_\mu \left(\bar{\gamma}^0-\bar{\gamma}^1\right)
\left\{\frac{i}{-(2\bp_0+\bp_1)\bp_1+i\epsilon}-2\pi n_F(|\bp_0|)\delta[(2\bp_0+\bp_1)\bp_1]\right\}.
\eqn{right}
\ee
The error in Ref.\ \cite{zhou}, responsible for the difference between LF and conventional amplitudes, occurs when this propagator is simplified as
\be
iS_{++}(\bp)\rightarrow 
  \bar{\gamma}^\mu \bp_\mu \left(\bar{\gamma}^0-\bar{\gamma}^1\right)\left\{
\frac{i}{-(2\bp_0+\bp_1)\bp_1+i\epsilon}-P^+ 2\pi n_F(|\bp_0|)\delta[(2\bp_0+\bp_1)\bp_1]\right\} \eqn{49prop}
\ee
(Eq.\ (49) of Ref.\ \cite{zhou}) where $P^+=-\frac{1}{2}\bgamma^1\bgamma^0$ 
is the projection operator on
the LF fermion dynamical degrees of freedom, $\psi_+=P^+\psi$, (see Eq.\ (25) in Ref.\ 
\cite{zhou}). The simplification indicated by \eq{49prop} is obtained by dropping the term where $-\pi n_F(|\bp_0|)$ is multiplied by
\begin{align}
2  \bar{\gamma}^\mu \bp_\mu &\left(\bar{\gamma}^0-\bar{\gamma}^1\right)(1-P^+)
\delta[(2\bp_0+\bp_1)\bp_1]\nn[2mm]
&=-\bgamma^0\bgamma^1(2\bp_0+\bp_1)\,\delta[(2\bp_0+\bp_1)\bp_1] \neq 0. \eqn{dropped}
\end{align}
That the term of \eq{dropped} has been dropped suggests that the mass shell 
$\delta$-function has been treated as 
$\delta[(2\bp_0+\bp_1)\bp_1]=\ds\frac{1}{|\bp_1|}\delta(2\bp_0+\bp_1)$ which is right only if
$\bp_1\neq 0$. In this way the 
part of the light-cone, $\bp_1=0$, is missing. In the
whole two-dimensional plane of variables $\bp_0$ and $\bp_1$, the correct formula is
\begin{eqnarray}
\delta[(2\bp_0+\bp_1)\bp_1]&=&\frac{1}{|\bp_1|} \delta(2\bp_0+\bp_1)+\frac{1}{2|\bp_0|} \delta(\bp_1).
\eqn{posneg}
\end{eqnarray}
In this respect, it is worth noting that although in the massive particle case ($m\ne 0$)
\be
\delta[(2\bp_0+\bp_1)\bp_1-m^2]=\frac{1}{|\bp_1|}\delta(2\bp_0+\bp_1-\frac{m^2}{\bp_1}),
\eqn{mposneg}
\ee
it would be wrong to conclude from this that 
$\delta[(2\bp_0+\bp_1)\bp_1-m^2] \rightarrow \ds\frac{1}{|\bp_1|}\delta(2\bp_0+\bp_1)$ in the zero mass limit.

Use of the propagator of  \eq{49prop} in the calculation of the thermal photon 
self-energy  leads to an amplitude (Eq.\ (53) of Ref.\ \cite{zhou}) whose nonanalyticity is due to just one $\delta$ function, $\delta(2\bp_0+\bp_1)$, which is only a part of the nonanalyticity obtained in the conventional approach where the term $\delta(\bp_1)$ is also present (but expressed in terms of variable $p^+=-\bp_1/\sqrt{2}$) \cite{silva} (here $\bp_0$ and $\bp_1$ are the total momentum components of the loop). It is a simple
exercise to restore the $\delta(\bp_1)$ term in the LF calculation of the
self-energy by keeping the last term in \eq{posneg} for the mass shell 
$\delta$-function. As in Ref.\ \cite{zhou}, let us consider the photon self-energy in
the real time formalism, namely, its $++$ component in the space of doubled degrees of freedom 
\begin{eqnarray}
\bPi^{\mu\nu}(\bp) = e^2
\int \frac{d^2\bk}{(2\pi)^2}\mbox{Tr}\,(\bgamma^0-\bgamma^1)\bgamma^\mu i S_{++}(\bk)(\bgamma^0-\bgamma^1)\bgamma^\nu i S_{++}(\bk+\bp) \eqn{Pi},
\end{eqnarray}
but with $S_{++}$ given by \eq{right} together with \eq{posneg} (rather than by \eq{49prop} as in Ref.\ \cite{zhou}). We note that the $\bgamma^\mu$ here are $2\times 2$ matrices \cite{silva}. Evaluating the temperature dependent part of \eq{Pi}, denoted by $\bPi^{\mu\nu(\beta)}(\bp)$, one finds  that
\be
\bPi^{00(\beta)}(\bp) = 2\bPi^{01(\beta)}(\bp) = 2\bPi^{10(\beta)}(\bp)
\ee
where
\begin{align}
\bPi^{01(\beta)}(\bp) &= -e^2\,\delta(2\bp_0+\bp_1)\int d\bk_1 \,\, \mbox{sgn}(\bk_1)\,\mbox{sgn}(\bk_1+\bp_1)\nn[2mm]
& \times \left[ n_F(|\bk_1|/2) + n_F(|\bk_1+\bp_1|/2) - 2 n_F(|\bk_1|/2)  n_F(|\bk_1+\bp_1|/2) \right],
\end{align}
which is in agreement with Ref.\ \cite{zhou}, but only because these particular tensor components
have momentum factors that nullify the contribution of the $\delta(\bp_1)$ term, even if it is retained in the calculation. However, such a momentum suppression of the $\delta(\bp_1)$ term does not occur for the tensor component $\bPi^{11(\beta)}(\bp)$. To see this, we write
\be
2\bPi^{11(\beta)}(\bp) = \bPi^{01(\beta)}(\bp) +\bPi^{(\beta)}(\bp)
\ee
so that
\begin{align}
\bPi^{(\beta)}(\bp) =  
e^2 \int \frac{d^2\bk}{2\pi}&\mbox{Tr}\,(2\bgamma^1-\bgamma^0) \kslash  \bgamma^1 
(\kslash +\pslash)
\left\{\frac{- i n_F(|\bk_0+\bp_0|)\delta[(\bk+\bp)^2]}{\bk^2+i\epsilon}\right. \nn[2mm]
&\left. +\, \frac{- i n_F(|\bk_0|)\delta(\bk^2)}
{(\bk+\bp)^2+i\epsilon} +2\pi n_F(|\bk_0+\bp_0|)\delta[(\bk+\bp)^2]
n_F(|\bk_0|)\delta(\bk^2) 
 \right\},
\end{align}
where  $\bk^2=-(2\bk_0+\bk_1)\bk_1$ and $(\bk+\bp)^2=-[2(\bk_0+\bp_0)+\bk_1+\bp_1](\bk_1+\bp_1)$.
Expressing the $\delta$ functions as a sum of two terms, as in \eq{posneg}, we see  that the trace,
$
\mbox{Tr}\,(2\bgamma^1-\bgamma^0) \kslash  \bgamma^1 
(\kslash +\pslash) =2(2\bk_0+\bk_1)\left[2(\bk_0+\bp_0)+\bk_1+\bp_1 \right]
$, nullifies some of these terms so that
\begin{align}
\bPi^{(\beta)}(\bp) &= e^2\int \frac{d^2\bk}{2\pi}
(2\bk_0+\bk_1)\left[2(\bk_0+\bp_0)+\bk_1+\bp_1\right]
\left\{\frac{- i n_F(|\bk_0+\bp_0|)\delta(\bk_1+\bp_1)}
{|\bk_0+\bp_0| [-(2\bk_0+\bk_1)\bk_1+i\epsilon]} \right. \nn[2mm]
& + \left. \frac{- i n_F(|\bk_0|)\delta(\bk_1)}
{|\bk_0| [-[2(\bk_0+\bp_0)+\bp_1]\bp_1+i\epsilon]}
+\frac{2\pi n_F(|\bk_0+\bp_0|)\delta(\bk_1+\bp_1)
n_F(|\bk_0|)\delta(\bk_1)}{2|\bk_0+\bp_0|\, |\bk_0|}
 \right\}.
\end{align}
After further simplification we obtain
\begin{align}
\bPi^{(\beta)}(\bp) &= - 2e^2\,\delta(\bp_1)\int d\bk_0 \,\, \mbox{sgn}(\bk_0)\,
\mbox{sgn}(\bk_0+\bp_0)\nn[2mm]
& \times \left[ n_F(|\bk_0|) + n_F(|\bk_0+\bp_0|) 
- 2 n_F(|\bk_0|)  n_F(|\bk_0+\bp_0|) \right].
\end{align}
It is now easy to check that these results for $\bPi^{00(\beta)}(\bp)$, $\bPi^{10(\beta)}(\bp)$, $\bPi^{01(\beta)}(\bp)$, and $\bPi^{11(\beta)}(\bp)$,  calculated using the LF formalism of Ref.\ \cite{zhou}, are identical to those calculated in the conventional case \cite{silva}.

Although the above analysis is straightforward, it needs to be recognized that the taking of the zero mass limit in $1+1$ dimensional LF field theory is well-known for its subtlety. Thus, although the right hand side of \eq{posneg} is the correct smooth massless limit of \eq{mposneg}, we would like to confirm, from the underlying fundamental quantum field theory of massless fermions, that it is just this smooth massless limit that is required to obtain the correct massless fermion propagator in the Schwinger model. For this purpose, we reexamine the above analysis from the point of view of the first consistent formulation of this problem worked out by McCartor some 15 years ago \cite{mccartor2}.

McCartor found that the correct way to quantize massless fields on the LF in $1+1$ dimensions is quite different from the way used to quantize massive fields (see below). From this fact one might be tempted to conclude that the massless propagator cannot be the smooth massless limit of the massive propagator; yet, as we shall see, it is. Following McCartor's quantization scheme, one obtains (see below) the following LF real time propagator for a massles fermion:
\be
D^L(p)=\frac{\pi}{2}\mbox{sgn}(p^-)\gamma^+\delta(p^+)
+\frac{i\gamma^-p^+}{p^2+i\epsilon}
-\pi\,\mbox{sgn}(p^0) \,n_F(|p_0|)\, [\gamma^+\delta(p^+)+\gamma^-\delta(p^-)]
\eqn{comprop}
\ee
where $p^\pm = \frac{1}{\sqrt{2}}(p^0\pm p^1)$ and $\gamma^\pm = \frac{1}{\sqrt{2}}(\gamma^0\pm \gamma^1)$. For easy comparison with the expressions of Ref.\ \cite{zhou}, we note that
$p^+=-\frac{1}{\sqrt{2}}\bp_1$, $p^-=\frac{1}{\sqrt{2}}(2\bp_0+\bp_1)$, $p^0=\bp_0$, and
$p^2 =-\bp_1(2\bp_0+\bp_1)$; also, the propagators of Ref.\ \cite{zhou}, and those denoted by $S_{++}$ above, are defined with an extra factor of $\gamma^0=\bgamma^0-\bgamma^1$ compared to those denoted by $D$'s above, below, and in Ref.\ \cite{blank}.
Some  interesting observations can now be made. Firstly, it is
evident that the temperature dependent part of the 
propagator, the last term of \eq{comprop}, is exactly the same as in the conventional
propagator, given by
\begin{eqnarray}\label{conv}
D(p)&=&p\llap/\left[\frac{i}{p^2+i\epsilon}-2\pi n_F(|p_0|)\delta(p^2)\right].
\end{eqnarray}
This is consistent with our observation in Ref.\ \cite{blank}, and mentioned above, that temperature is never the source of non-equivalence between LF and conventional theories.
Secondly, the LF propagator of \eq{comprop} coincides with
the smooth massless limit of the LF spinor propagator of Refs.\ \cite{blank} and \cite{beyer},
\be
D^{L}(p,m)=(p\llap/_{\mbox{\ssz on}}+m)\left[
\frac{i}{p^2-m^2+i\epsilon}-2\pi n_F(|p_0|)\delta(p^2-m^2)\right] \eqn{DLF}
\ee
where $p_{\mbox{\ssz on}}$ is the on mass shell momentum:
$p^-_{\mbox{\ssz on}}=m^2/2p^+$, $p^+_{\mbox{\ssz on}}=p^+$. Note, that in order to obtain this coincidence, the massless limit has to be taken in the mathematically correct way, which means that the 
$\delta$ function in \eq{DLF}, $\delta(p^2-m^2) = \delta[\bp_1(2\bp_0+\bp_1)+m^2]$, must reduce in the massless limit to a sum of two $\delta$ functions as given by \eq{posneg}.

Comparing now the propagator of \eq{right} with the ones denoted with $D$'s above, we see that it is identical with the conventional propagator given in \eq{conv}. Our finding that the propagator of \eq{right} gives photon self-energies that are identical with those calculated in conventional theory, is therefore not surprising. What is at issue, however, is the proper way to take the zero mass limit, and this is answered by the straightforward reduction of \eq{DLF} into \eq{comprop}, as just discussed.

To complete our analysis, we outline the derivation of  \eq{comprop}. The LF quantization of  massless fields in $1+1$ dimensions, formulated in Ref.\ \cite{mccartor2}, prescribes the following anticommutators of  fermion fields on the hyperplanes 
$x^+=0$ and $x^-=0$:
\begin{subequations}
\begin{align}
\{\psi_+(x),\psi^\dagger_+(0)\}_{x^+=0}&=\frac{1}{\sqrt{2}}P^+\delta(x^-)\\[2mm]
\{\psi_-(x),\psi^\dagger_-(0)\}_{x^-=0}&=\frac{1}{\sqrt{2}}P^-\delta(x^+)
\end{align}
\end{subequations}
where $P^\pm=\frac{1}{2}(1\pm\gamma^0\gamma^1)$ are the projection operators (the same $P^+$ operator was used previously in \eq{49prop}), and $\psi_\pm = P^\pm \psi$ are the dynamical ($+$) and non-dynamical ($-$) field components.
In combination with the free field equations 
$\partial\psi_+(x)/\partial x^+ =0$ and $\partial\psi_-(x)/\partial x^- =0$,
one obtains the anticommutator in the entire space
\begin{eqnarray}
\{\psi(x),\bar\psi(0)\}&=&\frac{1}{2}\left[\gamma^+\delta(x^+)+\gamma^-\delta(x^-)\right].
\end{eqnarray}
This commutator is just  the free LF spectral function whose Fourier transform is 
\begin{eqnarray}\label{comrho}
\rho^L_F(p)&=&\pi\left[\gamma^+\delta(p^+)+\gamma^-\delta(p^-)\right].
\end{eqnarray}
Now the spinor particle propagator, defined as the ensemble average
of the $x^+$-time ordered product, can be 
written in terms of the LF Lehmann representation as \cite{blank}
\begin{eqnarray}\label{spect}
D^L(p)&=&i\int \frac{dp'_0}{2\pi}
\frac{\rho^L(p_0',p^+)}{p_0-p_0'+i\epsilon}-
n_F(p_0)\rho^L(p).
\end{eqnarray}
In the free case, substituting $\rho^L(p)=\rho_F^L(p)$ as given by \eq{comrho}, one gets \eq{comprop}.
This completes our analysis and allows us to make a number of further observations.

Firstly, the above analysis shows explicitly that in LF thermal field theory of massless spinors in $1+1$ dimensions, both the dynamical and non-dynamical components get thermalized. This is in contradiction to the assertions made in Refs.\ \cite{zhou,das1,das2}. Although the  analysis presented here was for real time propagators, the same conclusion is reached when a similar analysis is made of imaginary time propagators \cite{new}.

Secondly, taking the smooth massless limit applies not just to propagators but to spectral functions as well.
For example, the free spectral function of \eq{comrho} is just the smooth massless limit of the free LF spinor spectral function of Ref.\ \cite{blank},
\be
\rho^L_F(p,m) = 2\pi\, \mbox{sgn}(p_0)\,(\ppslash+m)\delta(p^2-m^2).   \eqn{rhom}
\ee
In this regard, although the part of \eq{comrho} defined as $\rho^{L-}_F(p)\equiv\pi\gamma^-\delta(p^-)$ cannot be considered as the massless limit of \eq{rhom}, one could nevertheless consider it as a "prescription of quantization" and use it in \eq{spect}. One then obtains 
\be
D^{L}(p)=\frac{i\gamma^-p^+}{p^2+i\eta}
-n_F(|p_0|)\pi\epsilon(p^0)\gamma^-\delta(p^-)
\eqn{dasprop}
\ee
whose temperature dependent part is different from the temperature dependent part of the conventional propagator. This, of course, is just what has been argued in Refs.\  \cite{zhou,das1,das2}. To understand the origin of this nonequivalence we examine how such a propagator can arise from a field theoretic point of view. In this regard we note that the spectral function
$\rho^L_F(p)=\pi\gamma^-\delta(p^-)$ corresponds to quantization specified by anticommutators: 
\begin{eqnarray}
\{\psi_+(x),\psi^\dagger_+(0)\}_{x^+=0}&=&\frac{1}{\sqrt{2}}P^+\delta(x^-),\hspace{1cm}
\{\psi_-(x),\psi^\dagger_-(0)\}_{x^-=0}=0.
\end{eqnarray}
Just these commutators were in fact used for quantization prior to the publication of McCartor's  work, and describe physics which is not equivalent to the Schwinger model, as  explained in Ref.\ \cite{mccartor2}. In this light, one can see that the calculation of the temperature dependent part of the photon self-energy diagram, as given in Ref.\ \cite{zhou} and discussed in Ref.\ \cite{das2},
corresponds to a treatment which is not only not equivalent to the conventional 
approach at finite temperature, as claimed in Refs.\ \cite{zhou,das1,das2}, but  it is not equivalent even at zero temperature.

\end{document}